\documentstyle[A4,11pt,epsfig,twoside]{article}
\newcommand{\bea}{\begin{eqnarray}}
\newcommand{\eea}{\end{eqnarray}}
\newcommand{\be}{\begin{equation}}
\newcommand{\ee}{\end{equation}}

\begin{document}
\setcounter{page}{1}
\title{ELECTRO-WEAK FITS AT \sc{Clic}
       \thanks{In collaboration with M. Battaglia and D. Dominici}}
\author{STEFANIA DE CURTIS
%$^1$
        \thanks{e-mail address: decurtis@fi.infn.it}
\\
\\
        %$^1$
         {\it Istituto Nazionale di Fisica Nucleare,
        Sezione di Firenze, Italy}}
%\\
%        $^2$ {\it You Name University, Somewhere}
\date{}
\maketitle
\begin{abstract}
\noindent The aim of the future linear colliders is to extend the
sensitivity to new physics beyond the reach of the  {\sc Lhc}.
Several models predict the existence of new vector resonances in
the multi-TeV region. We review the existing limits on the masses
of these new  resonances from {\sc Lep/Slc} and {\sc Tevatron}
data and from the atomic parity violation measurements, in some
specific models. We study the potential of a multi-TeV $e^+e^-$
collider, such as {\sc Clic}, for the determination of their
properties and nature.
\end{abstract}
%***********************************************************************
\section{Introduction}
%***********************************************************************
\label{sec:intro} If new physics exists at the TeV scale it will be discovered
at the {\sc Lhc} and the precision measurements at a linear collider (LC) will
be crucial for revealing the character of the new phenomena. If the scale of
new physics is higher, than one of the most striking manifestation will come
from the sudden increase of the $e^+e^- \rightarrow f \bar f$ cross section
indicating the s-channel production of a new particle. New neutral vector
gauge bosons characterise several extensions of the Standard Model (SM) and
may represent the main phenomenology beyond 1~TeV. A first class consists of
models with extra gauge bosons such as a new neutral $Z'$ gauge boson. This is
common to both GUT-inspired $E_6$ models and to Left-Right (LR) symmetric
models.  Additional resonances are also introduced by recent string theories
in the form of Kaluza-Klein (KK) graviton and gauge boson excitations. We
discuss here a five dimensional extension of the Standard Model. Present
bounds are derived from precision electro-weak data and from the atomic parity
violation (APV) measurements. Beyond discovery, it will be essential to
accurately measure the mass, width, production and decay properties of the new
resonance to determine its nature and identify which kind of new physics it
manifests. This could be the task of a multi-TeV $e^+e^-$ collider such as
{\sc Clic}. Besides, {\sc Clic} will also probe  mass scales much beyond the
kinematic threshold for the production of the new vector gauge bosons. To
establish the sensitivity reach to indirect effects of new vectors as $Z'$
gauge bosons  and  KK excitations of the photon and of the $Z^0$ boson, we
consider cross sections, $\sigma_{f \bar f}$, and forward-backward asymmetries,
$A_{FB}^{f\bar f}$, for $\mu^+\mu^-$,  $b \bar b$ and $t \bar t$ at 1
TeV$<\sqrt{s}<$5 TeV.

%***********************************************************************
\section{New Vector Resonances:  $Z'$ and KK Excitations}
%***********************************************************************
\label{sec:Z'}

One of the best motivated extensions of the SM is given by the extra $U(1)$
models which predict an extra gauge boson $Z'$ associated to an additional
$U(1)$ symmetry. This extra symmetry, whose breaking scale is close to the
Fermi scale, is for example foreseen in some grand unified theories. Popular
models are parameterized by specific values of the angle $\theta_6$, which
defines the embedding of the extra $U(1)$ factor in $E_6$. The $\chi, \psi$
and $\eta$ models correspond to the values $\theta_6$=0, $\theta_6=\pi/2$ and
$\theta_6=-\tan^{-1}\sqrt{5/3}$ respectively. An extra neutral gauge boson is
also predicted by the Left-Right (LR) models in which the gauge group $SO(10)$
is broken down to the standard gauge group plus an additional $SU(2)_R$
factor. Finally, a useful reference is represented by the so-called sequential
standard model (SSM), which introduces an extra $Z'$ boson with the same
couplings of the SM $Z^0$ boson.\hfill\break There exist several constraints
on the properties of new neutral vector gauge bosons.  Direct searches for a
new $Z'$ boson  set lower limits on the masses from $\sigma(pp\to Z') B(Z'\to
ll)$. The 95\%~CL bounds on $M_{Z'}$ for  $\chi$, $\psi$, $\eta$, LR and SSM
models are 595, 590, 620, 630, 690 GeV respectively ({\sc Cdf} data
\cite{Abe:1997fd}). An extra $Z'$ naturally mixes with the SM $Z^0$ boson. The
present precision electro-weak data constrain the mixing angle, $\theta_M$,
within a few mrad \cite{Langacker:2001ij}. They also constrain the  $Z'$ mass.
From the average of the four {\sc Lep} experiments, for
 $\theta_M=0$, the  95\%~CL limit on $M_{Z'}$ for  $\chi$,
 $\psi$, $\eta$,  LR and  SSM models are
673, 481, 434, 804, 1787 GeV respectively \cite{kobel}. A third
class of constraints is derived from the atomic parity violation
(APV) data~\cite{Casalbuoni:1999yy}. In \cite{jhep} these bounds
are updated. The old analysis was based on the 1999 result of
weak charge $Q_W$ in the Cesium atomic parity experiment which
indicated a $\simeq 2.6 \sigma$ discrepancy w.r.t. the SM
prediction. A series of theoretical papers have since improved
the prediction of $Q_W$, by including the effect of the Breit
interaction among electrons and by refining the calculation of
radiative corrections \cite{APV}. The present value of $Q_W$
extracted from the Cesium data differs only $\simeq 0.8 \sigma$
from the SM prediction. Models involving extra neutral vector
bosons can modify the $Q_W$ value significantly. Assuming no
$Z^0-Z'$ mixing, we can evaluate the contribution to the weak
charge due to the direct exchange of the $Z'$ and derive
 bounds on $M_{Z'}$.
The  95\%~CL limit from APV on $M_{Z'}$ for the $\chi$,
 $\eta$,  LR and  SSM models are 627,
476, 665 and 1010 GeV respectively (no bounds can be set on the
$\psi$ model by the APV measurements). They are less stringent
than, or comparable to those from {\sc Lep} experiments. However,
as these bounds are very sensitive to the actual value of $Q_W$
and its uncertainties, further determinations may improve the
situation.\hfill\break
 The {\sc Lhc}  will push
the direct sensitivity to new vector gauge bosons beyond the TeV threshold.
With an integrated luminosity of 100~fb$^{-1}$, {\sc Atlas} and {\sc Cms} are
expected to observe signals from $Z'$ bosons for masses up to 4-5~TeV
depending on the specific model~\cite{Godfrey:2002tn}. A  future linear
$e^+e^-$ collider with c.o.m. energy below threshold can put lower bounds on
$M_{Z'}$ by measuring cross sections and asymmetries in the fermionic
channels. Both high luminosity and polarization of the electron and the
positron beams are effective in increasing the reach. For example the
sensitivity to the $\chi$ model at a 500 GeV LC with $L=.5$ ab$^{-1}$ will be
of 5 TeV by assuming $P_{e^+}$=.6 and $P_{e^-}$=.8 \cite{sabine}. Extra-$U(1)$
models can be accurately tested at a  linear  collider operating in the
multi-TeV region, such as {\sc Clic}. The observation of a $Z'$ signal is
granted but  the accuracy that can be reached in the study of its properties
depends on the quality of the accelerator beam energy spectrum and on the
detector response, including accelerator induced backgrounds.
 $M_{Z'}$,
$\Gamma_{Z'}/\Gamma_{Z^0}$ and $\sigma_{peak}$ can be extracted
by a resonance scan. The dilution of the analysing power due to
the beam energy spread is appreciable. Still, the relative
statistical accuracies can be better than 10$^{-4}$ on the mass
and $5 \times 10^{-3}$ on the width~\cite{scan}.

Theories of quantum gravity have considered the existence of extra-dimensions
for achieving the unification of gravity at a scale close to that of
electro-weak symmetry breaking. String theories have recently suggested that
the SM could live on a $3+\delta$ brane with $\delta$ compactified large
dimensions while gravity lives on the entire ten dimensional bulk. The
corresponding models lead to new signatures for future colliders ranging from
Kaluza-Klein (KK) excitations of the gravitons to KK excitations of the SM
gauge fields with masses in the TeV range. We consider here a five-dimensional
extension of the SM with fermions on the boundary. This predicts KK
excitations of the SM gauge bosons with fermion couplings $\sqrt{2}$ larger
compared to those of the SM \cite{Pomarol:1998sd}. Masses of the KK
excitations of $W$, $Z^0$ and $\gamma$ are given by $M_n \simeq nM$, for large
value of the fifth dimension compactification scale, $M$.
 The  KK excitations of $W^{\pm}$, $Z^0$ and $\gamma$ mix
with the SM gauge bosons. This mixing is parameterized by the
angle $\beta$.\hfill\break
 Indirect limits from electro-weak
measurements  are derived by considering the modifications in the electro-weak
observables at the $Z^0$ peak and at low energy \cite{ewlimits}. An update of
the   lower bounds on the compactification scale  derived from a recent
determination of the $\epsilon$ parameters and from the latest APV results is
given in~\cite{jhep}. The $95\%$~CL limits from  high energy experiments range
from $M>2.5$ TeV for  $\sin \beta =0$ to $M>4.5$ TeV for $\sin\beta=1$. The
ones from APV turn out to be  significantly below the corresponding high energy
bounds.\hfill\break
 Non observation of deviations in lepton pair
production at {\sc Lhc} can set limits on the compactification
scale $M$. For example by considering an effective luminosity of
5 fb$^{-1}$ one gets a bound of
$M=6.7$~TeV~\cite{Antoniadis:1999bq}. At {\sc Clic}, the lowest
excitations $Z^{(1)}$ and $\gamma^{(1)}$ could be directly
produced. Results for the $\mu^+\mu^-$ cross sections and
forward-backward asymmetries at the Born level and after folding
the effects of the  beam spectrum are given in~\cite{jhep}.

\section{Indirect
Sensitivity to $Z'$ and KK Excitations at {\sc Clic}}

\begin{figure}[t]
\begin{center}
\begin{tabular}{c c c}
\hspace{-1cm}\includegraphics[width=0.36\textwidth,height=0.4\textwidth]{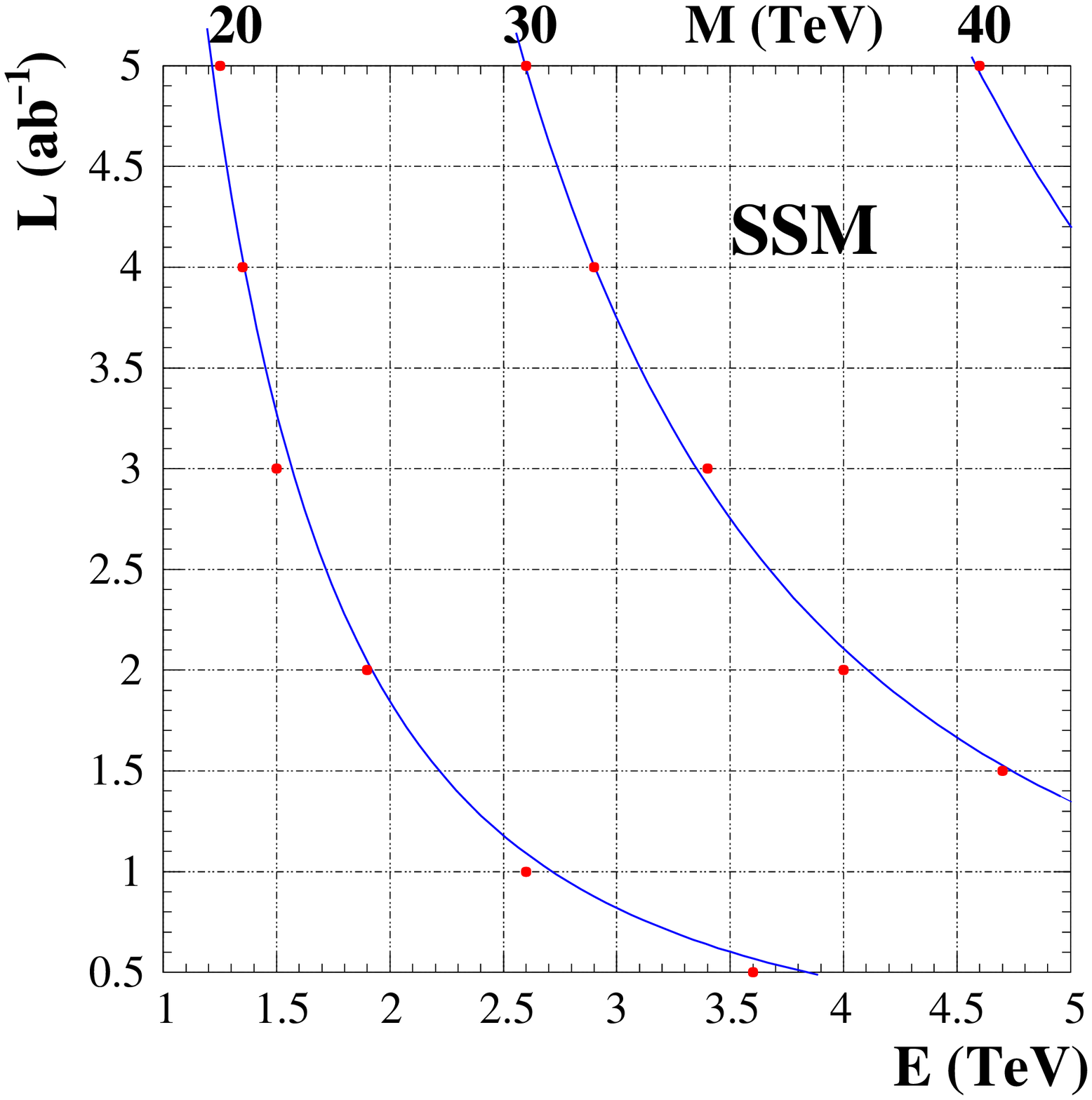}&
\includegraphics[width=0.33\textwidth,height=0.4\textwidth]{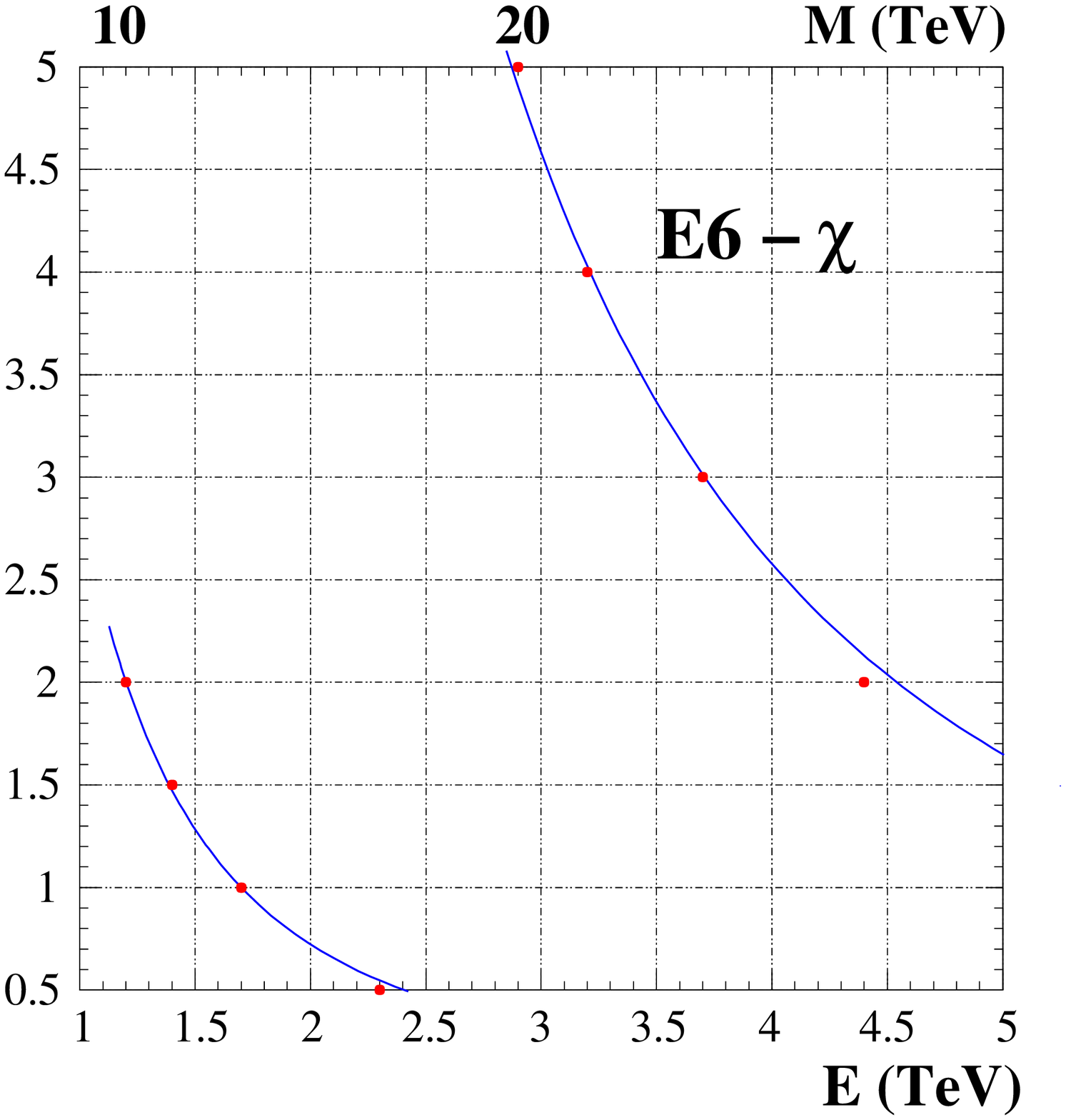}&
\includegraphics[width=0.33\textwidth,height=0.4\textwidth]{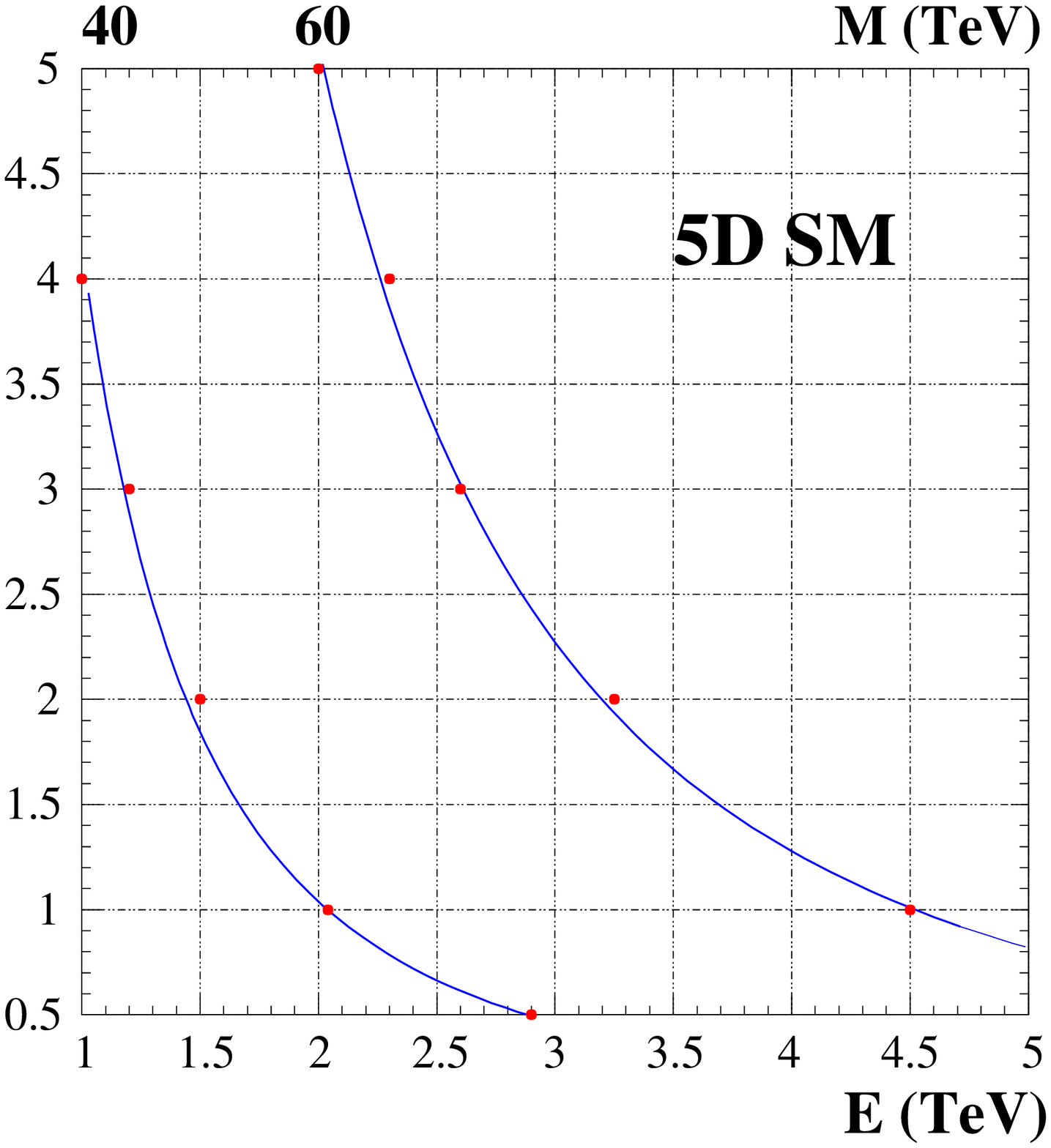}\\
\end{tabular}
\caption{\it{The 95\%~CL sensitivity contours in the ${L}$ vs.
$\sqrt{s}$ plane for different values of $M_{Z'}$ in the SSM
model (left), in the $E_6$-$\chi$ model (center) and in the 5D~SM.
The points represent the results of the analysis, while curves
show the behaviour expected from the scaling $M_{Z'} \propto (s
L)^{1/4}$.} \label{fig:zp}}
\end{center}
\end{figure}

Precision electro-weak measurements performed in multi-TeV $e^+e^-$ collisions
can push the mass scale sensitivity for scenarios beyond the SM beyond the
10~TeV frontier. We consider here the $\mu^+\mu^-$, $b \bar{b}$ and $t
\bar{t}$ production cross sections $\sigma_{f \bar{f}}$ and forward-backward
asymmetries $A_{FB}^{f \bar{f}}$.\hfill\break As a general property, at a LC
the indirect sensitivity to the mass of a new vector resonance such a $Z'$,
can be parameterized in terms of the available integrated luminosity $L$, and
c.o.m energy, $\sqrt{s}$. In fact a scaling law for large $M_{Z'}$ can be
obtained by considering the effect of the $Z'-\gamma$ interference in the
cross section. For $s<< M_{Z'}^2$ and assuming that the uncertainties $\delta
\sigma$ are statistically dominated, we get that the range of mass values
giving a significant difference from the SM prediction can be derived from:
$|\sigma^{SM} - \sigma^{SM+Z'}|/\delta \sigma \propto \sqrt{sL}/M^2_{Z'}
> \sqrt{\Delta \chi^2}$. This means that
the sensitivity to the $Z'$ mass scales as: $M_{Z'} \propto (s L)^{1/4}$. This
relationship shows that there is a direct possible trade-off  between
$\sqrt{s}$ and ${L}$, which should be taken into account when optimizing the
parameters of a high energy $e^+e^-$ linear collider.\hfill\break
 At the {\sc Clic}
design c.o.m. energies, the relevant $e^+e^- \rightarrow f \bar f$ cross
sections are significantly reduced and the experimental conditions at the
interaction region need to be taken into account in validating the accuracies
on electro-weak observables. Since the two-fermion cross section is of the
order of only 10~fb, it is imperative to operate the collider at high
luminosity. This can be achieved only in a regime where beam-beam effects are
important and primary $e^+e^-$ collisions are accompanied by several $\gamma
\gamma \rightarrow {\mathrm{hadrons}}$ interactions. Being mostly confined in
the forward regions, this $\gamma \gamma$ background reduces the polar angle
acceptance for quark flavour tagging and dilutes the jet charge separation
using jet charge techniques. These experimental conditions require efficient
and robust algorithms to ensure sensitivity to flavour-specific $f \bar f$
production. The statistical accuracies for the determination of $\sigma_{f
\bar f}$ and $A_{FB}^{f \bar f}$ have been studied using a realistic
simulation \cite{Battaglia:2000iw,laura}.  The results are summarized in terms
of the relative statistical accuracies on the electro-weak observables
obtained for 1~ab$^{-1}$ of {\sc Clic} data at $\sqrt{s}$ = 3~TeV, including
the effect of $\gamma \gamma \rightarrow {\mathrm{hadrons}}$
background~\cite{jhep}: $\delta\sigma_{\mu^+\mu^-}/\sigma_{\mu^+\mu^-}=\pm
0.010$, $\delta\sigma_{b \bar b}/\sigma_{b \bar b}=\pm 0.012$,$\delta\sigma_{t
\bar t}/\sigma_{t \bar t}=\pm 0.014$, $\delta
A_{FB}^{\mu\mu}/A_{FB}^{\mu\mu}=\pm 0.018$, $\delta A_{FB}^{b \bar b
}/A_{FB}^{b \bar b}=\pm 0.055$, $\delta A_{FB}^{t \bar t }/A_{FB}^{t \bar
t}=\pm 0.040$. The $\sigma_{f \bar f}$ and $A_{FB}^{f \bar f}$ ($f =
\mu,~b,~t$) values have been computed, for 1~TeV $< \sqrt{s} <$ 5~TeV, both in
the SM and including the corrections due to the presence of a $E_6-Z'$ boson
with 10~TeV $< M_{Z'} <$ 40~TeV. Predictions have been obtained by
implementing these models in the {\sc Comphep} program~\cite{comphep}.  The
sensitivity has been defined as the largest $Z'$ mass giving a deviation of
the actual values of the observables from their SM predictions corresponding
to a SM probability of less than 5\%. The SM probability has been defined as
the minimum of the global probability computed for all the observables and
that for each of them, taken independently. This sensitivity has been
determined, as a function of the $\sqrt{s}$ energy and integrated luminosity
$L$ and compared to the previous scaling law. Results for the SSM and for the
$\chi$ model are summarized in Figure~\ref{fig:zp}. For the $\eta$ model the
sensitivity is lower: for example to reach a sensitivity of $M_{Z'}$=20~TeV,
more than 10~ab$^{-1}$ of data at $\sqrt{s}$=5~TeV would be
necessary.\hfill\break
\hspace{0.4cm}In the case of the 5D~SM, we have included only the
effect of the exchange of the first KK excitations $Z^{(1)}$ and
$\gamma^{(1)}$, neglecting that of the remaining excitations of
the towers, which give only small corrections. The scaling law
for the limit on the compactification scale  $M$ can be obtained
by considering the interference of the two new nearly degenerate
gauge bosons with the SM photon in the cross section and taking
the $s<<M^2$ limit. The result is again $M \propto (s L)^{1/4}$.
\begin{figure}[t]
\begin{center}
\epsfig{file=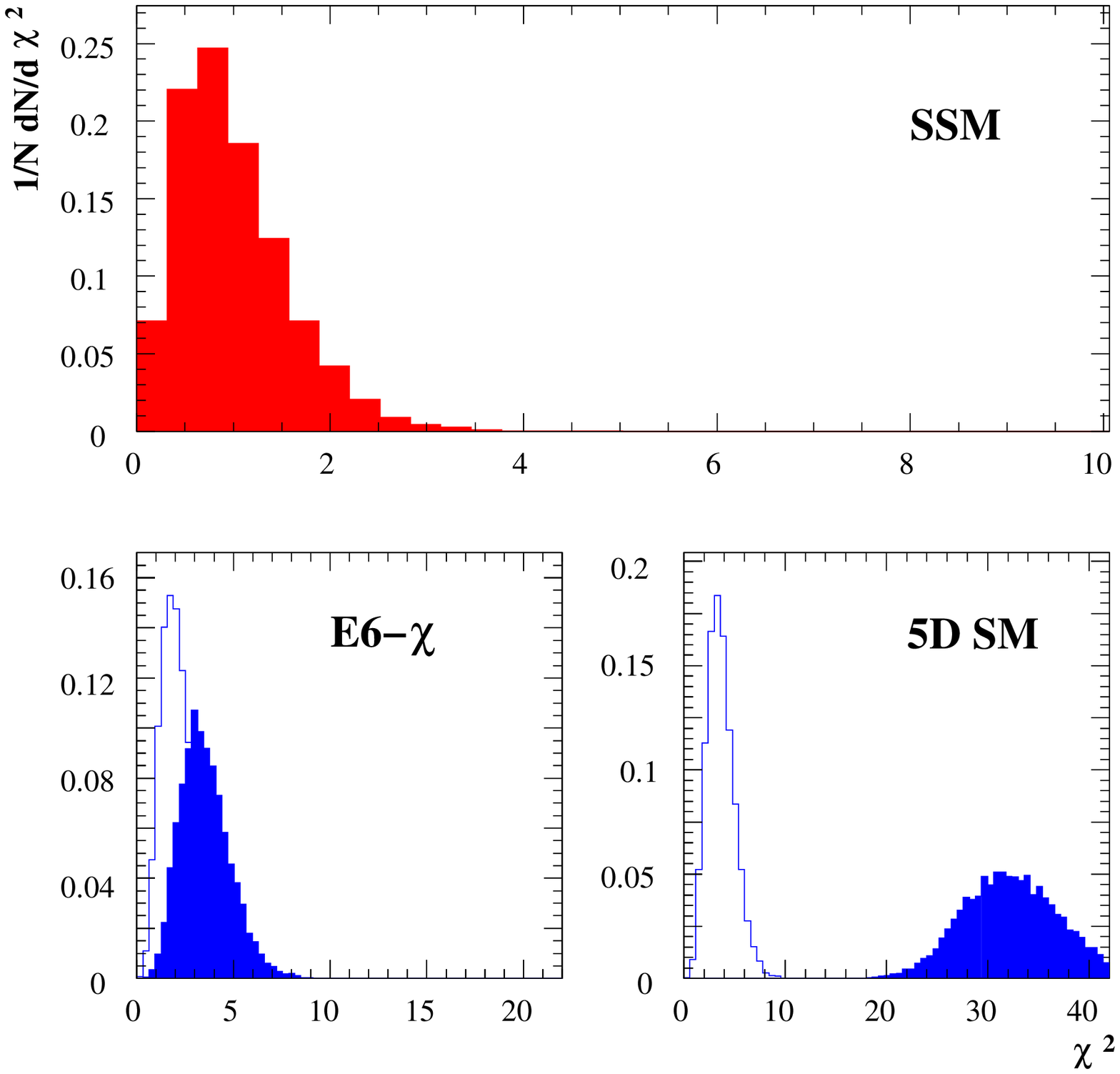,width=12.cm,height=9.cm,clip} \caption{\it $\chi^2$
distributions obtained for a set of pseudo-experiments (${L}$ = 1~ab$^{-1}$ of
{\sc Clic} data at $\sqrt{s}$~=~3~TeV) where the SSM is realized with a $Z'$
mass of 20~TeV (upper plot). The corresponding distributions for the
$E_6$-$\chi$ and 5D~SM for $M$=20~TeV (full histograms) and for $M$=40~TeV are
shown in the lower panels.  } \label{fig:sep}
\end{center}\end{figure} The analysis closely follows that for
the $Z'$ boson discussed above. In Figure~ \ref{fig:zp} we give
the sensitivity contours as a function of $\sqrt{s}$  for
different values of $M$. We conclude that the sensitivity
achievable on the compactification scale $M$ for an integrated
luminosity of 1~ab$^{-1}$ in $e^+e^-$ collisions at $\sqrt{s}$ =
3-5~TeV is of the order of 40-60~TeV. Results for a similar
analysis, including all electro-weak observables, are discussed
in~\cite{Rizzo:2001gk}.

An important issue concerns the ability to probe the models, once a
significant discrepancy from the SM predictions would be observed. Since the
model parameters and the mass scale are {\it a priori} arbitrary, an
unambiguous identification of the scenario realized is difficult. However,
some informations can be extracted by testing the compatibility of different
models while varying the mass scale. Suppose that a particular model is
realized in nature. By integrating the $\chi^2$ distribution for a different
model, we can evaluate the confidence level at which the two models are
distinguishable. Figure~\ref{fig:sep} shows an example of such tests. Taking
$M$=20~TeV, ${L}$ = 1~ab$^{-1}$ of {\sc Clic} data at $\sqrt{s}$~=~3~TeV could
distinguish the SSM model from the $E_6$-$\chi$ model at the 86\% CL and from
the 5D~SM at the 99\% CL.  For a mass scale of 40~TeV, $L$ = 3~ab$^{-1}$ of
{\sc Clic} data at $\sqrt{s}$~=~5~TeV, the corresponding confidence levels
become 91\% and 99\% respectively. Further sensitivity to the nature of the
gauge bosons could be obtained by studying the polarized forward-backward
asymmetry $A_{FB}^{pol}$ and the left-right asymmetry $A_{LR}$ colliding
polarized beams.

\section{Conclusions}

 Accuracies achievable for the
determination of the fundamental properties of new neutral vector resonances
are discussed for different classes of models, using realistic assumptions for
the experimental conditions at {\sc Clic}. Even beyond the kinematical reach
for s-channel production, a multi-TeV $e^+e^-$ collider could probe the
existence of new vector resonances up to scales of several tens of TeV by
studying the unpolarised electro-weak observables. Some information regarding
the nature of these new resonances could still be gained and further
sensitivity would be provided by the use of polarised beams.

\end{document}